**Title:** Energy-efficient ultrafast SOT-MRAMs based on low-resistivity spin Hall metal $Au_{0.25}Pt_{0.75}$

Authors: Lijun Zhu*, Lujun Zhu, Shengjie Shi, D. C. Ralph, and R. A. Buhrman

Dr. Lijun Zhu, Dr. S. Shi, Prof. R. A. Buhrman
Cornell University,
Ithaca, New York 14850, USA
Email: lz442@cornell.edu

Prof. Lujun Zhu
College of Physics and Information Technology,
Shaanxi Normal University,
Xi'an 710062, China

Prof. D. C. Ralph
Cornell University,
Ithaca, New York 14850, USA
Kavli Institute at Cornell,
Ithaca, New York 14850, USA

**Abstract**: Many key electronic technologies (e.g., large-scale computing, machine learning, and superconducting electronics) require new memories that are fast, reliable, energy-efficient, and of low-impedance at the same time, which has remained a challenge. Non-volatile magnetoresistive random access memories (MRAMs) driven by spin-orbit torques (SOTs) have promise to be faster and more energy-efficient than conventional semiconductor and spin-transfer-torque magnetic memories. This work reports that the spin Hall effect of low-resistivity $Au_{0.25}Pt_{0.75}$ thin films enables ultrafast antidamping-torque switching of SOT-MRAM devices for current pulse widths as short as 200 ps. If combined with industrial-quality lithography and already-demonstrated interfacial engineering, our results show that an optimized MRAM cell based on $Au_{0.25}Pt_{0.75}$ can have energy-efficient, ultrafast, and reliable switching, e.g. a write energy of < 1 fJ (< 50 fJ) for write error rate of 50% (<$10^{-5}$) for 1 ns pulses. The antidamping torque switching of the $Au_{0.25}Pt_{0.75}$ devices is 10 times faster than expected from a rigid macrospin model, most likely because of the fast micromagnetics due to the enhanced non-uniformity within the free layer. These results demonstrate the feasibility of $Au_{0.25}Pt_{0.75}$-based SOT-MRAMs as a candidate for ultrafast, reliable, energy-efficient, low-impedance, and unlimited-endurance memory.

**1. Introduction**

Many key electronic technologies, e.g., large-scale computing, machine learning, and superconducting electronics, would benefit from the development of new fast, non-volatile, and energy-efficient memories.[1-3] While the conventional non-volatile 2-terminal spin-transfer-torque (STT) magnetoresistive random access memory (MRAM) is attractive for its good scalability and high thermal stability[4,5] during fast sub-ns write,[6] the required high write current density can exert severe stress on the magnetic tunnel junction (MTJ) and induce wear-out and breakdown of the MTJ barrier,[7] leading ultimately to degradation of the memory cell. Meanwhile, the shared read/write path can lead to write upon read errors. An alternative, 3-terminal spin-orbit-torque (SOT) MRAM [8,9] has the potential to mitigate these issues. In a SOT-MRAM, the spin current generated by the spin Hall effect (SHE)[10-13] of a heavy metal layer switches the magnetic free layer of a MTJ (see Figure 1a). The non-volatile SOT-MRAMs can have long data retention, zero standby power, and fast and reliable write.[7,14-17] SOT-MRAMs based on a spin Hall metal that combines a giant spin Hall ratio ($\theta_{SH}$) with a relatively low resistivity ($\rho_{xx}$) can also have unlimited endurance due to the suppression of Joule heating induced bursting and migration of the write line [2] as well as low values of write impedance that is compatible with superconducting circuits in cryogenic computing systems.[1]

Recent harmonic response measurements [18,19] have indicated that the $Au_{0.25}Pt_{0.75}$ alloy can be a particularly compelling spin Hall metal for high-performance SOT-MRAM application due to the combination of a relatively low resistivity ($\rho_{xx} \approx 80$ μΩ cm) with a giant antidamping SOT efficiency ($\xi_{DL}^j \approx 0.3$-0.35). In this work, we show that prototype SOT-MRAM devices based on $Au_{0.25}Pt_{0.75}$ can achieve highly energy-efficient, ultrafast (down to 200 ps), and reliable switching. The antidamping torque switching of the $Au_{0.25}Pt_{0.75}$ devices is 10 times faster than expected from a rigid macrospin model, most likely because of the enhanced non-uniformity within the free layer.

## 2. Results and Discussion

### 2.1 Characterization of the SOT-MRAM Devices

As shown in Figure 1b-d, the SOT-MRAM devices were lithographically patterned from sputter-deposited multilayer stacks consisting of Si/SiO$_2$/Ta 1/Au$_{0.25}$Pt$_{0.75}$ 5/Hf 0.5/ Fe$_{0.6}$Co$_{0.2}$B$_{0.2}$ 1.4/Hf 0.1/MgO 1.6/Fe$_{0.6}$Co$_{0.2}$B$_{0.2}$ 4/Pt 3/Ru 4 (numbers are layer thicknesses in nm), and were post annealed at 240 °C for 1 hour. The 0.1 nm-thick Hf spacer inserted at the Fe$_{0.6}$Co$_{0.2}$B$_{0.2}$/MgO interface is to reduce the demagnetization field ($4\pi M_{eff}$) of the free layer by enhancing the interfacial perpendicular magnetic anisotropy.[15] Low value of $4\pi M_{eff}$ reduces the critical current for antidamping switching.[20,21] Ferromagnetic resonance measurements on the unpatterned films yield magnetic damping $\alpha = 0.027 \pm 0.001$ and $4\pi M_{eff} = 0.460 \pm 0.003$ T for the magnetic free layer (see Figure S1). As determined by vibrating sample magnetometry, the saturation magnetization ($M_s$) of the Fe$_{0.6}$Co$_{0.2}$B$_{0.2}$ layers is 1240 emu/cm$^3$.[22,23] All measurements were performed at room temperature.

Figure 1b is a top-view scanning electron microscopy (SEM) image of one of the SOT-MRAM devices, showing an elliptical (190×45 nm$^2$) MTJ with the long axis transverse to the spin Hall channel and thus to the write-current flow. The spin Hall channel is 300 nm wide in the center where the pillar is located (Figure 1b and Figure S2) and 1.2 μm long as measured by the cross-sectional transmission electron microscopy (TEM) image (Figure 1c). For the sake of fabrication simplicity, we make electrical contact to the two ends of the channel using two micron-size trapezoid-shaped MTJ pillars (marked as "via" regions in Figure 1b,c, see Supporting Information for more details about the dimensions of the "via" and an estimate that each "via" contributes an effective series resistance of approximately 200 Ω). The channel resistance ($R_{ch}$), including the "via" resistance and the series resistance from the two "vias" was 850 Ω, much lower than that of previously reported SOT-MRAMs from our laboratory (2.5-4.2 kΩ for W,[24] Ta,[9] Pt/Hf multilayers [23]) due to the reduced channel length and the relatively low $\rho_{xx}$ of Au$_{0.25}$Pt$_{0.75}$.[18]

Figure 1e,f show the major and minor magnetic switching loops of a representative device for in-plane magnetic fields applied along the long axis of the MTJ pillar. The major loop indicates a coercivity ($H_c$) of 450 Oe for the 4 nm-thick Fe$_{0.6}$Co$_{0.2}$B$_{0.2}$ reference layer due to the shape anisotropy of the elliptical MTJ pillar. The minor loop is artificially centered after subtraction of the dipole field ($H_{dipole} \approx 150$ Oe) from the 4 nm reference layer. The minor loop indicates an $H_c$ of 15 Oe for the 1.4 nm free layer and a tunnel magnetoresistance ratio of 20% for the MTJ.

### 2.2 Direct Current Switching

Figure 1g shows the characteristic switching behavior of the same SOT-MRAM device as the dc write current in the spin Hall channel is ramped quasi-statically (with an in-plane field equal to -$H_{dipole}$ applied along the long axis of the MTJ pillar to compensate the dipole field from the reference layer). The MTJs show abrupt switching at write currents of ≈75 μA. Since thermal fluctuations assist the reversal of a nanoscale MTJ during slow current ramps,[21,25,26] we carried out ramp rate measurements (Figure 1h). Within the macrospin model, the switching current $I_c$ should scale with the ramp rate ($\dot{I}$) following [20]

$$I_c = I_{c0}\left(1 + \frac{1}{\Delta}\ln\frac{t_0\Delta|\dot{I}|}{|I_{c0}|}\right), \qquad (1)$$

where $I_{c0}$ is the critical switching current in absence of thermal fluctuations, $\Delta$ the stability factor equal to the magnetic energy barrier for reversal between the P and AP states normalized by the thermal energy $k_BT$, and $t_0$ the thermal attempt time which we assume to be 1 ns. By fitting the data to Equation (1), we obtain $|I_{c0}| = 312 \pm 11$ μA for P→AP and $356 \pm 14$ μA for AP→P switching, and $\Delta \approx 28 \pm 2$. These results were consistently reproduced by other devices. For practical application, $\Delta$ can be increased significantly even for sub-100-nm devices by optimizing the shape anisotropy during pillar etching process and by introducing tensile strain anisotropy.[27] Considering a parallel resistor model,[23] the current shunted into the $Fe_{0.6}Co_{0.2}B_{0.2}$ free layer and Hf spacers ($\rho_{Pt/Hf} \approx 80$ μΩ cm, $\rho_{FeCoB} \approx \rho_{Hf} \approx 130$ μΩ cm) can be estimated to be $\approx 0.2 I_{c0}$. The critical switching density in the $Au_{0.25}Pt_{0.75}$ spin Hall channel is, therefore, $j_{c0} = (1.86 \pm 0.08) \times 10^7$ A/cm$^2$ for P→AP switching $(2.12 \pm 0.07) \times 10^7$ A/cm$^2$ for AP→P switching, which are a factor of 2 lower than that for devices with pure Pt channels [14] in which $\alpha$ and $4\pi M_{eff}$ are even smaller (Table 1).

Within the simple macrospin model, $j_{c0}$ for antidamping torque switching of an in-plane magnetized MTJ can be estimated by [21,26]

$$j_{c0} = (2e/\hbar) A \mu_0 M_s t \alpha (H_c + 4\pi M_{eff}/2)/\xi_{DL}^j, \quad (2)$$

where the factor $A \approx \exp(-d_{Hf}/\lambda_{s,Hf})$ denotes the attenuation of spin current by the Hf spacer layer in between the spin Hall channel and the magnetic free layer of the MTJ. With the Hf thickness $d_{Hf} = 0.5$ nm and the spin diffusion length $\lambda_{s,Hf} = 0.9 \pm 0.2$ nm,[28] we determine $A \approx 0.57 \pm 0.08$ for the SOT-MRAMs. From Equation (2) we estimate $\xi_{DL}^j$ to be $0.30 \pm 0.07$, which is consistent with our previous harmonic response measurements on $Au_{0.25}Pt_{0.75}$/Co bilayers without a Hf spacer ($\xi_{DL}^j \approx 0.3$-$0.35$).[18] As compared in Table I, the value of $\xi_{DL}^j \approx 0.30$ is significantly higher than those previously obtained in W devices ($\xi_{DL}^j \approx -0.20$),[24] Pt devices ($\xi_{DL}^j \approx 0.12$),[14] and $Pt_{0.85}Hf_{0.15}$ devices ($\xi_{DL}^j \approx 0.23$)[17] when the spin current attenuation by the Hf spacer layers ($A \leq 1$) is taken into account (note that $A$ was assumed to be unity in previous reports [14,15,17,23] when calculating $\xi_{DL}^j$). The SOT efficiency is similar to [Pt 0.6/Hf 0.2]$_6$/Pt 0.6 multilayers ($\xi_{DL}^j \approx 0.29$, $\rho_{xx} \approx 140$ μΩ cm [23]), but the lower-resistivity $Au_{0.25}Pt_{0.75}$ ($\rho_{xx} \approx 85$ μΩ cm) is more favorable for applications that require unlimited endurance [2] and low device impedance.[1] $Au_{0.25}Pt_{0.75}$ is also thermally stable as indicated by the constant $\rho_{xx}$ and $\theta_{SH}$ upon annealing to 400 °C.[19]

## 2.3. Ultrafast and reliable pulse current switching

We characterized the performance of the MRAMs in the short-pulse regime using a measurement method similar to that described in Refs. [15,17]. Figure 2a shows the switching phase diagram in the pulse width ($\tau$) regime of 0.2-6 ns for the two cases AP→P and P→AP, respectively. Each data point is the statistical switching probability result of 1000 switching attempts. In determining the current values plotted in Figure 2, we have taken into account the impedance discontinuity between the 50 Ω cable and the MRAM channel,[17,29] so that the currents quoted denote the real pulse magnitudes within the channel. We find that the low $\rho_{xx}$ and giant $\theta_{SH}$ of $Au_{0.25}Pt_{0.75}$ [18] allow the MRAM device to be switched many millions of times in the sub-ns pulse regime with no indication of degradation in the MgO barrier or the spin Hall channel.[2] For 200 ps pulses, the write current ($I$) for 50% switching probability are 3 mA (AP→P) and 3.27 mA (P→AP), and for 400 ps pulses both are $\approx 2$ mA. The write energy ($E_{write} = I^2 R_{ch} \tau$) of the $Au_{0.25}Pt_{0.75}$ device at the current corresponding to 50% switching probability is plotted as a function of $\tau$ in Figure S3. $E_{write}$ is as low as 1 pJ, 1.4 pJ, and 2 pJ for 1 ns, 400 ps, and 200 ps switching, respectively. This is encouraging as the values of $\alpha$, $4\pi M_{eff}$, and channel dimensions could all be reduced further by additional optimization so that the write current and energy can be decreased significantly (see below).

While as discussed below it is apparent that our devices do not reverse as a rigid single domain when driven by strong SOT pulses, we can still parameterize a time scale ($\tau_0$) characteristic of the switching process from fits of the 50% switching probability points to the macrospin model prediction [21]

$$I = I_\infty (1 + \tau_0/\tau), \quad (3)$$

where $I_\infty$ denotes the critical switching current at infinite pulse width. As shown in Figure 2b, we find $\tau_0 = 1.52 \pm 0.02$ ns and $I_\infty = 0.441 \pm 0.005$ mA for P→AP switching and $\tau_0 = 0.86 \pm 0.01$ ns and $I_\infty = 0.617 \pm 0.005$ mA for AP→P switching. The write current density ($j_\infty$) of 2.1 (3.0)$\times 10^7$ A/cm$^2$ for P→AP (AP→P) switching is higher than the zero-temperature dc switching current density from the ramp rate. This difference represents initial evidence that the SOT-induced magnetic reversal in the short-pulse regime is not well-described by macrospin dynamics, because $I_\infty$ and $I_{c0}$ should be close in the case that a macrospin moment is excited by the antidamping

spin torque.[21]

It has been a consistent observation that, in the short pulse regime, the spin-torque switching of the in-plane magnetized SOT-MRAMs ($\tau_0 < 2$ ns)[2,14-16] and metallic spin valves ($\tau_0 \approx 1$ ns)[26,30] are much faster than the prediction of the macrospin model. We find that the $Au_{0.25}Pt_{0.75}$ devices are more than a fact of 10 faster and more energy-efficient than that expected for a rigid macrospin. As indicated by Bedau *et al.*,[26] the characteristic time for antidamping torque switching of a macrospin nanomagnet can be estimated as

$$\tau_0 \approx (4\pi M_{eff}\alpha\gamma)^{-1} \qquad (4)$$

where $\gamma$ is the gyromagnetic ratio. With the experimental values of $\alpha$ and $4\pi M_{eff}$ of the actual $Au_{0.25}Pt_{0.75}$ devices, Eq. (4) yields $\tau_0 \approx 18$ ns for a macrospin reversal process, much slower than our measurements. Because the switching can be quite fast, our $Au_{0.25}Pt_{0.75}$ devices are much more energy efficient than expected by a rigid macrospin model in short pulse regime. For example, for the pulse width of 200 ps (1 ns), the required switching current for the $Au_{0.25}Pt_{0.75}$ device is $6I_\infty$ ($2I_\infty$), markedly smaller than $90I_\infty$ ($20I_\infty$) predicted by the macrospin simulation (see Figure 2b).

Understanding the switching mechanism of the in-plane devices are the key to develop ultrafast memory for technological applications. Here we attribute the observed ultrafast switching mainly to the enhanced the non-uniform micromagnetic dynamics within the free layer of our devices. As have been suggested by previous efforts,[31-34] the antidamping torque switching of the in-plane magnetized free layer is achieved via a fast evolution of non-uniform micromagnetic dynamics rather than via a coherent macrospin reversal within the free layer. The magnetic non-uniformity of the free layer should enhance the micromagnetic dynamics and speed up the switching. As schematically shown in Figure 2d, the SOT-MRAMs fabricated in our group have substantial tapering (see Ref. [23] for TEM imaging of the tapering) in the MTJ free layer that results from the ion-milling process, which should result in spatially non-uniform SOTs and dipole field within the magnetic free layer. There is also strong interfacial Dzyaloshinskii-Moriya interaction (DMI)[35] and magnetic roughness (variations of thickness and interfacial magnetic anisotropy field)[22] at the Pt (alloy)/FM interfaces, which should enhance the magnetic non-uniformity.[36] This explains the fact that our Pt (alloy) based in-plane SOT-MRAMs ($\tau_0 \approx 1$ ns) with enhanced tapering and DMI are typically faster than the W devices [2] and the TaB devices ($\tau_0 \approx 3.3$-$3.4$ ns as fitted in Figure S4)[37,38] where both the tapering of free layer and the interfacial DMI are relatively weak.[39] Another factor that could assist the evolution of non-uniform dynamics is the current-induced effective transverse field ($H_{eff}$), which is the sum of the fieldlike SOT effective field ($H_{FL}$) and Oersted field ($H_{Oe}$) in the SOT-MRAM geometry. Early micromagnetic simulations [14,31] show that $H_{eff}$, if parallel to the spin polarization of the spin Hall current, can speed up the non-uniform dynamics and thus the switching of the free layer. As indicated by the dc switching phase diagrams ($H_c$ vs $I$) in Figure 2e, for our $Au_{0.25}Pt_{0.75}$ devices a positive (negative) charge current induced $H_{eff}$ reduce $H_c$ for AP→P (P→AP) switching. This indicates that $H_{eff}$ is parallel to the spin polarization of the spin current from the $Au_{0.25}Pt_{0.75}$ channel and therefore could play an assisting role in antidamping torque switching of the free layer. However, the magnetic non-uniformity of the free layer turns out to be more critical than $H_{eff}$ in determining the switching speed. As we compare in Figure 2e,f, the device discussed above ($H_{eff} / I \approx 115$ Oe/mA, $I_\infty$=0.44 mA, denoted as Device A) is not as fast as a control device (denoted as Device B, the stack is $Au_{0.25}Pt_{0.75}$ 4/FeCoB 1.6/MgO 2/FeCoB 4) that has two times smaller effective field ($H_{eff}/ I \approx 24$ Oe/mA, $I_\infty$=1.09 mA). This difference in the $\tau_0$ values is likely suggestive of a less significant non-uniformity in the free layer of Device A than in the Device B. A thin Hf layer has been consistently found to reduce the interfacial spin-orbit coupling (thus most likely interfacial DMI) at heavy metal/ferromagnetic interfaces.[19] Finally, $H_{eff}$ in the short strong pulse region can exceed the value of $H_c$ ($H_c$= 15 Oe for Device A, 40 Oe for Device A), which might be reminiscent of a switching driven directly by $H_{eff}$. However, the fact that Device B is faster than Device A reaffirms that it is the antidamping torque rather than $H_{eff}$ which dominates the switching process. This conclusion is also supported by the rhombehedral shape of the bistable region (P/AP) in the dc phase diagram (Figure 2e). We speculate that magnetization switching by an effective field that is collinear with the magnetization seems to be slow because the excitation of the dynamics likely requires accumulation of thermal fluctuation for a nonzero initial torque. In any case, the very short characteristic switching time of our devices is a very positive observation for application and motivates further study on the switching mechanisms in 3-terminal SOT-MRAMs.

For memory, SOT reversal must be both fast and highly reliable. While the limited-statistics switching probability for pulse current and duration sweeps (for instance, 1000 events in Figure 2a) are routinely used to report the existence of high-speed switching,[4,16,17] they do not convey the statistics of the write error rate (WER)—information that is critical for applications. We have tested the reliability of our $Au_{0.25}Pt_{0.75}$ devices by measuring WER statistics during up to $10^5$ switching attempts. As shown in Figure 2c, WERs for the pulse duration of 1 ns scale down quickly as the write current increases, extrapolation of which indicates WERs of < $10^{-5}$ at 4 mA ($2.2×10^8$ A/cm$^2$) for both the P→AP and the AP→P switching.

Further decreases in the required write currents of $Au_{0.25}Pt_{0.75}$ SOT-MRAM devices can be expected from straightforward additional optimization of the stack materials and device dimensions. Interface engineering has already been demonstrated to significantly reduce both $α$ and $4πM_{eff}$ of the $Fe_{0.6}Co_{0.2}B_{0.2}$ free layer.[19,20,22] For example, our optimized MRAMs based on a spin Hall channel of [Pt 0.6/Hf 0.2]$_6$ multilayers [23] or W [14] achieved $α$ of ≈0.011 and $4πM_{eff}$ of ≈ 0.2 T, both of which are more than a factor of 2 less compared to our present $Au_{0.25}Pt_{0.75}$ devices. As shown in Fig. S5 we have found that a Pt 0.5/Hf 0.25 bilayer can effectively suppress $α$ contributed by the interfacial spin-orbit coupling via spin memory loss [19] and two-magnon scattering,[22] despite that the 0.5 nm-thick Hf single-layer spacer that was inserted at the bottom of the 1.4 nm $Fe_{0.6}Co_{0.2}B_{0.2}$ free layer appears ineffective in doing so probably because Hf does not wet Au surface and forms a discontinuous layer. The write current and the write power can be further reduced by additional factors of 3 and >18 by using more aggressive industry-level lithography to narrow the spin Hall channel from 300 nm to below 100 nm [27,37] and to shorten it from 1.2 μm to 200 nm,[7] and by replacing the magnetic stacks in the "via" regions (see Figure 1c and Figure S2) with highly conductive materials. These approaches in combination would lower the zero temperature switching current (density) of $Au_{0.25}Pt_{0.75}$ SOT-MRAM devices to $I_{c0}$ < 30 μA ($j_{c0}$ < $4.8×10^6$ A/cm$^2$) and $I_∞$ < 60 μA ($j_∞$ < $9.6×10^6$ A/cm$^2$). The write energy for 50% switching probability for 1 ns will be < 1fJ. Assuming similar switching dynamics, for reliable switching with 1 ns pulse and the WER of <$10^{-5}$, the write current (density) would scale to < 0.35 mA ($5.6×10^7$ A/cm$^2$) and the write energy to < 50 fJ.

## 3. Conclusion

We have demonstrated that $Au_{0.25}Pt_{0.75}$ is a particularly compelling spin Hall metal that can enable very energy-efficient and ultrafast switching of in-plane-magnetized SOT-MRAMs due to the combination of a giant spin Hall effect ($ξ_{DL}^j$ ≈ 0.30) and a low resistivity ($ρ_{xx}$ ≈ 80 μΩ cm). We have demonstrated switching of prototypical SOT-MRAM structures with 50% probability using $I$ ≈ 3 mA and $E_{write}$ = 2 pJ for 200 ps current pulses, and write error rates < $10^{-5}$ at $I$ = 4 mA and $E_{write}$ = 14 pJ for 1 ns pulses. We extrapolate that further reductions of $α$ and $4πM_{eff}$ of the free layer (as already demonstrated in other SOT-MRAM structures) along with improved lithography to reduce the dimension of the spin Hall channel can enable 1 ns switching of SOT-MRAM devices with write energy < 50 fJ and WER of < $10^{-5}$. The relatively low channel resistance due to the low $ρ_{xx}$ of $Au_{0.25}Pt_{0.75}$ is beneficial for decreasing write energies, achieving unlimited endurance, and also for matching the impedance of superconducting circuits in cryogenic computation systems. We find that the current-induced SOT switches the $Au_{0.25}Pt_{0.75}$-based MRAMs much faster than expected from a rigid macrospin model, most likely due to the rapid micromagnetics within the free layer that is enhanced by the spatial non-uniformities in the free-layer magnetization that may be induced by DMI, interfacial magnetic roughness, and/or tapering in the MTJ free layer. If combined with the strain and voltage gating architectures proposed in the industry,[27,37,38] the $Au_{0.25}Pt_{0.75}$-based in-plane SOT-MRAMs can be also very dense. The non-volatile MRAM also have long data retention and zero standby power. These results indicate that the $Au_{0.25}Pt_{0.75}$-based SOT-MRAM is a good candidate for ultrafast, energy-efficient, low-impedance, unlimited-endurance memory for large scale computing systems, machine-learning systems, and superconducting electronics.

**Experimental section**

*Sample growth and device fabrication*: All of the samples are sputter deposited at room temperature with an argon pressure of 2 mTorr and a base pressure of ~ $1×10^{-8}$ Torr. A highly resistive oxidized Si substrate ($ρ_{xx}$ > $10^{10}$ μΩ cm) was used to avoid current shunting into the substrate during the direct current and the pulse current switching measurements. The 1 nm Ta layer at the bottom was used to improve the smoothness and adhesion of the $Au_{0.25}Pt_{0.75}$. The top bilayers of Pt 3 nm/Ru 4 nm were used to protect the multilayers during device fabrication. The multilayer samples are patterned into 3-terminal MRAM devices schematically shown in **Figure S2** with a three-step procedure. First, we defined the spin Hall channel using DUV lithography (ASML) and ion beam etching and measured the channel size to be 300×600 nm$^2$ by atomic force microscopy (Veeco Icon). We then defined the elliptical MTJ nanopillars with different aspect ratios and μm-size "via" pillars (as vertical connector between the bottom channel to top contact) onto the spin Hall channel with e-beam lithography (JEOL JBX-6300FS) and ion beam etching, and isolated the pillars with 80 nm thick SiO$_2$ deposited by an e-beam evaporator. Finally, contacts of Ti 5 nm/Pt 50 nm were sputter-deposited on the top of the MTJ pillars and "via" pillars for electrical measurements.

*Measurements*: For the dc switching measurements of the MRAM devices, a lock-in amplifier was used to read the differential resistance of the magnetic tunnel junctions with a 0.1 V oscillatory voltage applied onto MTJ pillars series-connected to 10 MΩ resistor (read current ≈ 1 µA). A Keithley 2400 source-meter was used to source write current into the spin Hall channels. For the short pulse measurement, the pulse was generated using a picosecond pulse generator and the MTJ resistance was measured with a NI-DAQ (voltmeter) and a Keithley 2450 (current source). A vibrating sample magnetometer was used to determine the sample magnetization. Flip-chip ferromagnetic resonance was used to determine the magnetic damping constant and the effective demagnetization field of the $Fe_{0.6}Co_{0.2}B_{0.2}$ free layer on large-area unpattern chips by sweeping an in-plane magnetic field at each fixed microwave frequency (see **Figure S1**). The MRAM devices were characterized by cross-sectional scanning transmission electron microscopy (TEM) imaging in a spherical-aberration-corrected (Cs-corrected) 300-kV FEI Titan G2 microscope.

**Acknowledgements**
This work was supported in part by the Office of Naval Research (N00014-15-1-2449) and by the NSF MRSEC program (DMR-1719875) through the Cornell Center for Materials Research. The devices were fabricated, in part, at the Cornell NanoScale Facility, an NNCI member supported by NSF Grant No. ECCS-1542081. The TEM measurements performed at Shaanxi Normal University were supported by the Science and Technology Program of Shaanxi Province (Grant No. 2019JQ-433) and the Fundamental Research Funds for the Central Universities (Grant No. GK201903024).

**Supporting information**:
Supporting Information is available from the Wiley Online Library or from the author.

**Conflict of Interest**
The authors declare no conflict of interest.

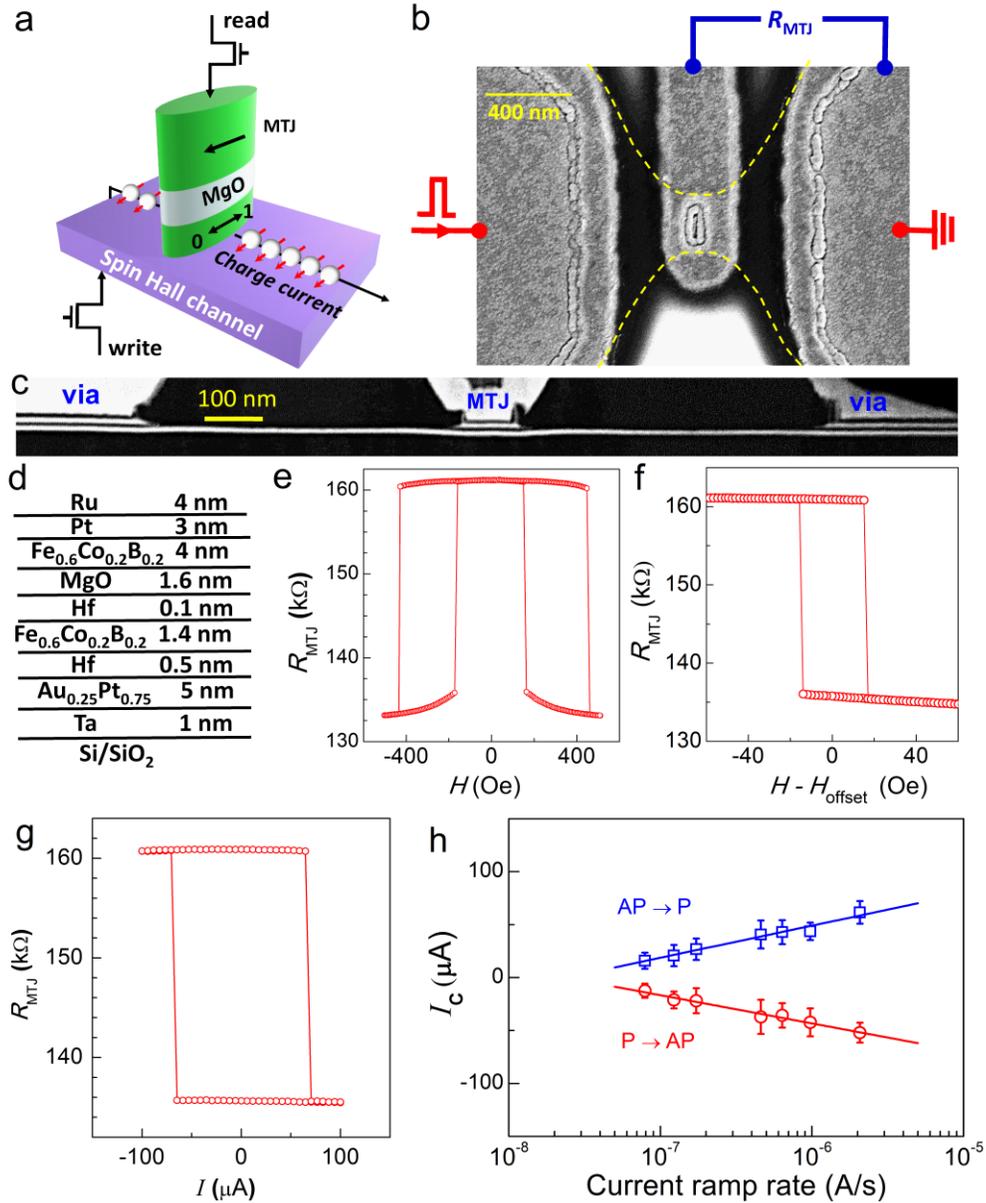

Figure 1. a) Schematic of a SOT-MRAM with an in-plane MTJ and a spin Hall channel. b) Top-view SEM image, c) Cross-sectional TEM image (dark field), d) Sample stack, e) Magnetic major loop, and f) Magnetic minor loop of the SOT-MRAM devices. g) DC switching loop, h) DC switching currents vs current ramp rate for P→AP (red circles) and AP→P (blue squares) switching as a function of current ramp rate. In b), the two yellow dashed lines indicates the area of the spin Hall channel; in h) the solid lines represents the best fit of the data to Equation (1).
 
 
 

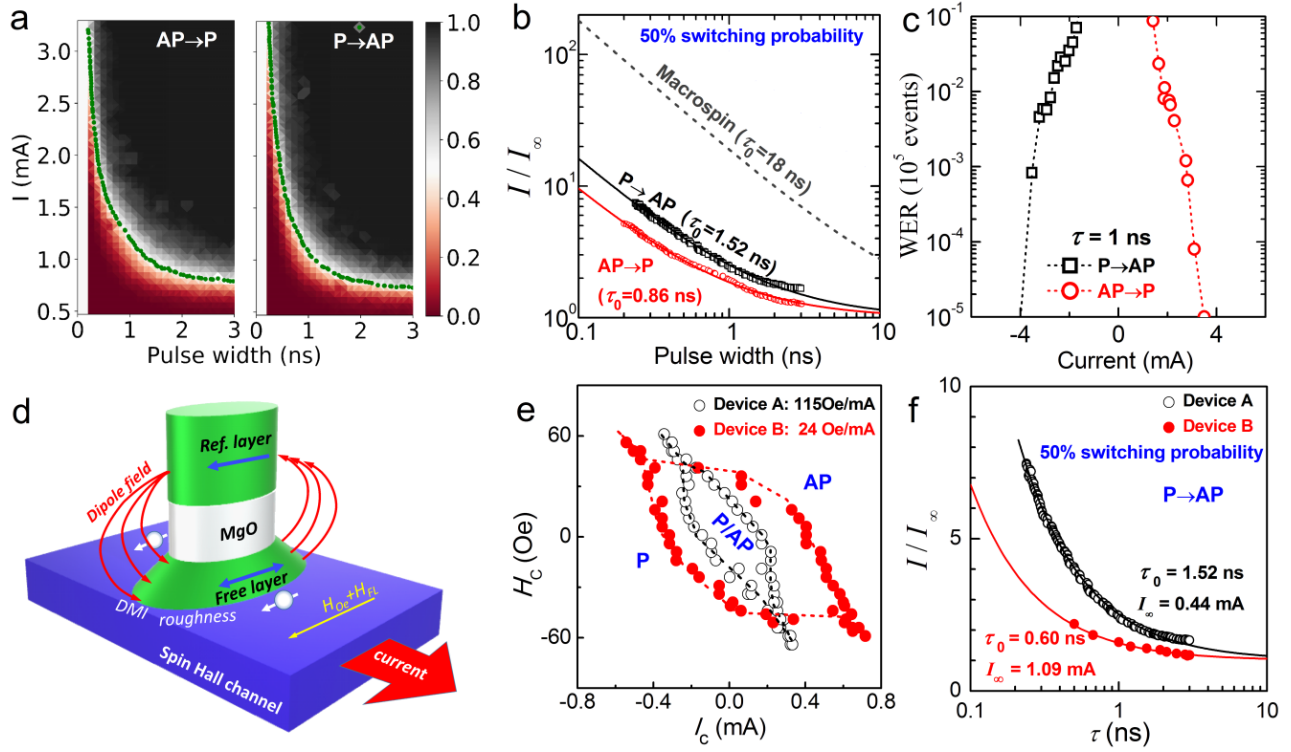

Figure 2. Fast pulse switching of the $Au_{0.25}Pt_{0.75}$-based MRAM device. (a) Pulse switching phase diagrams (the color scale represents the switching probability for 1000 events), (b) Rescaled write current for AP→P (red) and for P→AP switching (black) plotted as a function of pulse width (50% switching probability), (c) The write error rates (WERs) with 1 ns pulse ($10^5$ events) plotted as a function of write current, and (d) Schematic depict of non-uniformities within the free layer. (e) DC phase diagrams and (f) Rescaled write current (P→AP switching, 50% switching probability) for two devices with different strength of current-induced effective field ($H_{eff}/I \approx 115$ Oe/mA for Device A and 24 Oe/mA for Device B). In (a), the green dots indicate the 50% switching probability points; in (b), the dashed gray lines denotes the macrospin reversal with a critical switching time of 18 ns as calculated using Equation (4); in (b) and (f) the solid lines represent the best fits to Equation (3) of the 50% switching probability points.

Table 1. Comparison of the parameters of the SOT-MRAM devices calculated using Equation (2), indicating the effectiveness of $Au_{0.25}Pt_{0.75}$ in generating dampinglike spin-orbit torque. The factor $A \approx \exp(-d_{Hf}/\lambda_{s,Hf})$ denotes the attenuation of spin current by the Hf spacer layer (the thickness $d_{Hf}$ and the spin diffusion length $\lambda_{s,Hf}$ in between the spin Hall channel and the magnetic free layer of the magnetic tunnel junction. $R_{ch}$ is the resistance of the spin Hall channel, $j_{c0}$ critical dc switching current density, $\alpha$ magnetic damping, $4\pi M_{eff}$ the demagnetization field, $\xi_{DL}^j$ the dampinglike spin-orbit torque efficiency per unit current density.

| Structure | $A$ | $R_{ch}$ (kΩ) | $j_{c0}$ ($10^7$ A/cm$^2$) | $\alpha$ | $4\pi M_{eff}$ (T) | $\xi_{DL}^j$ | Refs |
|---|---|---|---|---|---|---|---|
| $Au_{0.25}Pt_{0.75}$ 5/Hf 0.5/$Fe_{0.6}Co_{0.2}B_{0.2}$ 1.4/Hf 0.1 | 0.57±0.08 | 0.85 | 2.0 | 0.027 | 0.460 | 0.30 | This work |
| [Pt 0.6/Hf 0.2]$_5$/Pt 0.6/$Fe_{0.6}Co_{0.2}B_{0.2}$ 1.6 | 1 | 4.3 | 1.0 | 0.017 | 0.553 | 0.29 | [23] |
| [Pt 0.6/Hf 0.2]$_6$/Pt 0.6/Hf 0.25/$Fe_{0.6}Co_{0.2}B_{0.2}$ 1.6/Hf 0.1 | 0.76±0.05 | 3.8 | 0.36 | 0.011 | 0.197 | 0.23 | [23] |
| W 4.4/Hf 0.25/ $Fe_{0.6}Co_{0.2}B_{0.2}$ 1.8/Hf 0.1 | 0.76±0.05 | 3.6 | 0.54 | 0.012 | 0.211 | -0.20 | [15] |
| $Pt_{0.85}Hf_{0.15}$ 6/Hf 0.7/ $Fe_{0.6}Co_{0.2}B_{0.2}$ 1.4 | 0.46±0.09 | 2.5 | 1.4 | 0.017 | 0.362 | 0.23 | [17] |
| Pt 5/Hf 0.7/$Fe_{0.6}Co_{0.2}B_{0.2}$ 1.6 | 0.46±0.09 | 1.05 | 4.0 | 0.018 | 0.4165 | 0.12 | [14] |
| Ta 6.2/$Fe_{0.4}Co_{0.4}B_{0.2}$ 1.6 | 1 | 3 | 3.7 | 0.021 | 0.76 | -0.12 | [9] |

**Supporting Information for**

**Energy-efficient ultrafast SOT-MRAMs based on low-resistivity spin Hall metal Au$_{0.25}$Pt$_{0.75}$**

Lijun Zhu[1*], Lujun Zhu[2], Shengjie Shi[1], D. C. Ralph[1,3], and R. A. Buhrman[1]

1. Cornell University, Ithaca, New York 14850, USA
2. College of Physics and Information Technology, Shaanxi Normal University, Xi'an 710062, China
3. Kavli Institute at Cornell, Ithaca, New York 14850, USA

* lz442@cornell.edu

Section 1. Ferromagnetic resonance measurement

Section 2. Write impedance of the SOT-MRAMs

Section 3. Calculated write power at 50% switching probability

Section 4. Literature data on the pulse switching of SOT-MRAMs

Section 5. Interfacial engineering of the damping

**Section 1. Ferromagnetic resonance measurement**

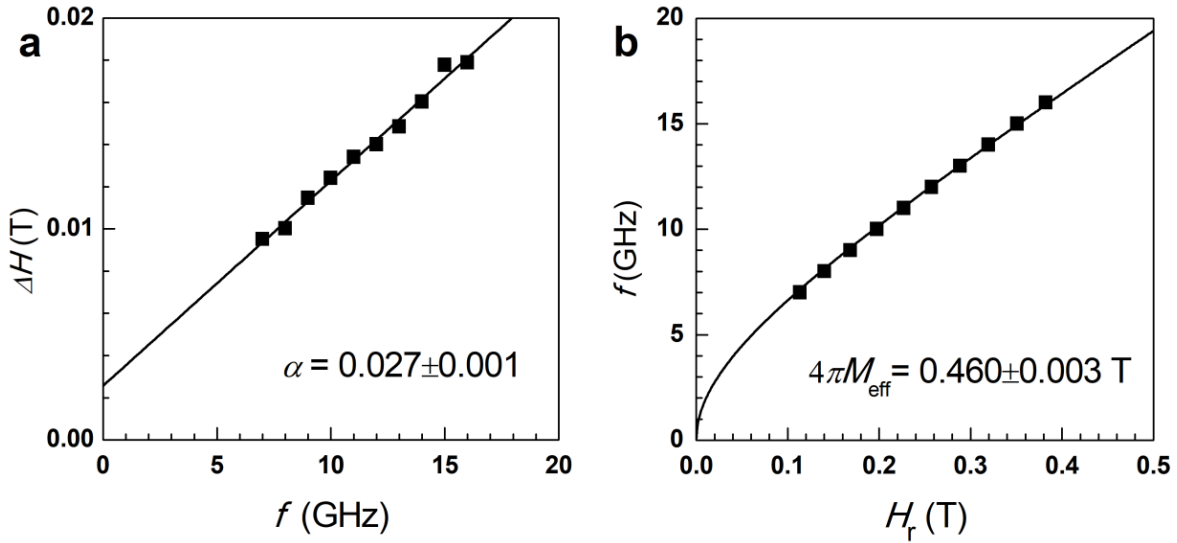

Figure S1. Ferromagnetic resonance measurement on the free layer of Device A. a) FMR linewidth $\Delta H$ vs the resonance frequency $f$, b) frequency vs FMR resonance field $H_r$. The solid lines in (a) and (b) represent the best fits to $\Delta H = \Delta H_0 + (2\pi/\gamma)\alpha f$, and $f = (\gamma/2\pi)\sqrt{H_r(H_r + 4\pi M_{\text{eff}})}$, respectively. $\Delta H_0$ and $\gamma$ are the inhomogeneous broadening of the FMR linewidth and the gyromagnetic ratio, respectively.

**Section 2. Write impedance of the SOT-MRAMs**

    The measured write impedance of our SOT-MRAM devices includes the resistance of the channel and the two "vias". The width of the spin Hall channel varies gradually from ≈10 μm (at two ends, see Figure S2) to 300 nm (at the center, see Figure 1b). Due to the spreading effect, the current flow is not uniform in the channel, especially at the two thick ends. As shown in Figure S2, the "vias" are trapezoid-shaped with the width narrowing gradually from 8 μm at the end far from the center of the channel to 1 μm close to the center of the channel. If we assume a uniform current flow in the "vias" during write of the device, the total resistance of the "vias" can be estimated to be ≈170 Ω by comparing the areas of the vias (15 $\mu m^2$) and the MTJ pillar (190×45 $nm^2$). However, because of the spreading effect, the current is mostly flowing at the narrow end of vias rather than flowing uniformly in the "via" region, which makes the electrically effective area of each "via" much less than 15 $\mu m^2$. A control Pt-based MRAM device with a similar device dimensions shows that the measured "channel resistance" is reduced by 400 Ω when the magnetic stack in "via" regions were etched away and filled with highly conductive Ti/Pt. Therefore, we can conclude that the resistance contribution of the two "via" regions should be approximately 400 Ω for our devices.

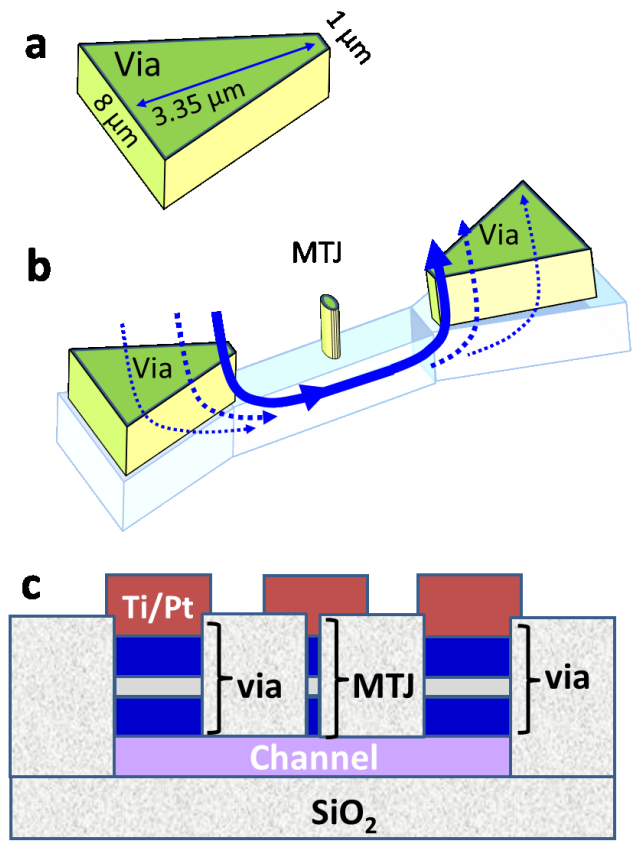

Figure S2. Schematic of the 3-terminal MRAM device. a) Dimensions of the "via" region; b) Spreading effect of the current flow; c) Side-view of the MRAM devices, indicating that the two μm-size "via" pillars are consist of the same magnetic stack as the nanopillar of the magnetic tunnel junction (MTJ) and contribute to the measured channel resistance of the device.

**Section 3. Calculated write power at 50% switching probability**

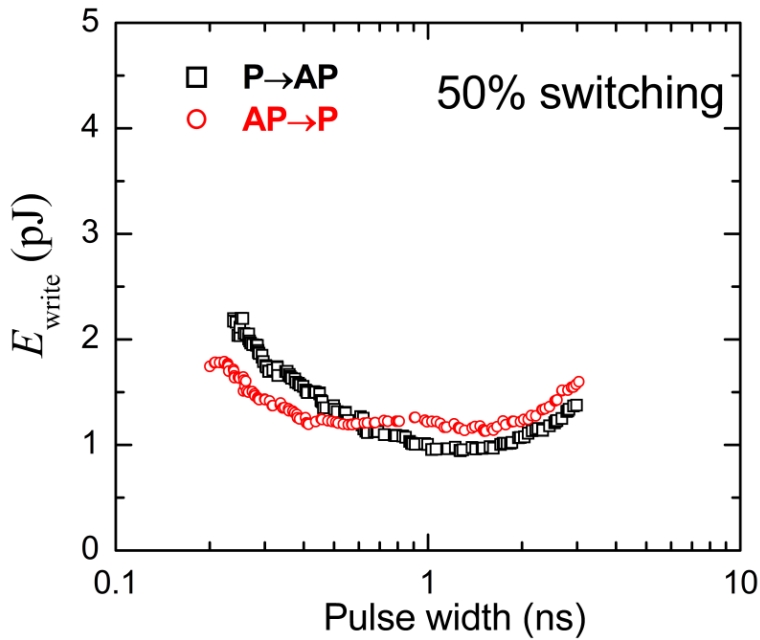

Figure S3. Write energy (50% switching probability) of a typical SOT-MRAM device based on $Au_{0.25}Pt_{0.75}$.

**Section 4. Literature data on the pulse switching of SOT-MRAMs**

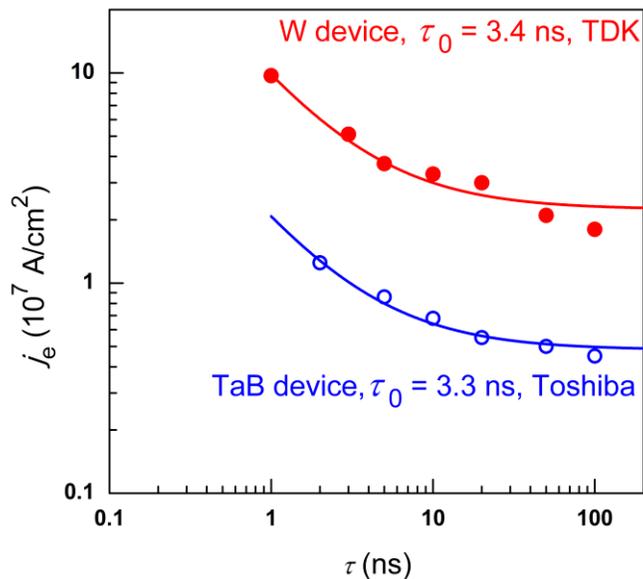

Figure S4. Write pulse width dependence of switching current density for 50% probability for in-plane magnetized SOT-MRAM based on a spin Hall channel of W (red dots, ref. 2) and TaB (blue circles, ref. 37). Fitting the data to Equation (3) in the maintext yields the critical switching time ($\tau_0$) of 3.4 ns for the W device and 3.3 ns for the TaB device.

## Section 5. Interfacial engineering of the damping

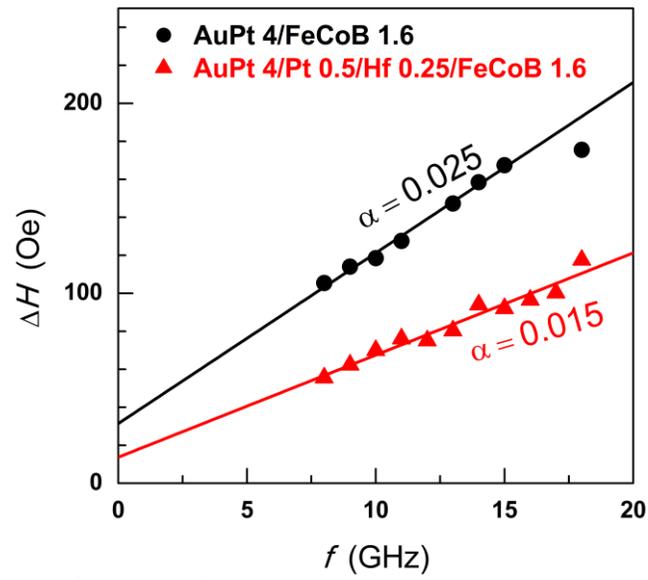

Figure S5. The FMR linewidth vs frequency for AuPt 4/FeCoB 1.6 and AuPt 4/Pt 0.5/Hf 0.25/FeCoB 1.6, indicating a substantial reduction of damping due to the insertion of the Pt 0.5/Hf 0.25 bilayer spacer.